\documentclass[superscriptaddress,pre,showpacs,twocolumn]{revtex4-1}
\usepackage{amssymb,amsmath,amsthm,color}
\usepackage{graphicx,amssymb,color}

\newcommand{\beq}{\begin{equation}}
\newcommand{\eeq}{\end{equation}}
\newcommand{\reff}[1]{(\ref{#1})}
\newcommand{\dif} {{\rm d}}

\begin{document}
\title{Active temperature and velocity correlations produced by a swimmer suspension}
\author{C. Parra-Rojas} 
\affiliation{Departamento de F\'isica, Facultad de Ciencias F\'isicas y 
Matem\'aticas, Universidad de Chile, Casilla 487-3, Santiago, Chile}

\author{R. Soto}
\affiliation{Departamento de F\'isica, Facultad de Ciencias F\'isicas y 
Matem\'aticas, Universidad de Chile, Casilla 487-3, Santiago, Chile}
\affiliation{
Rudolf Peierls Centre for Theoretical Physics, University of Oxford, Oxford
OX1 3NP, UK}
\date{\today}

\begin{abstract}
The agitation produced in a fluid by a suspension of micro-swimmers in the low Reynolds number limit is studied. In this limit, swimmers are modeled as force dipoles all with equal strength.
The agitation is characterized by the active temperature defined, as in kinetic theory, as the mean square velocity, and by the equal-time spatial correlations. 
Considering the phase in which the swimmers are homogeneously and isotropically distributed in the fluid, it is shown that 
the active temperature and velocity correlations depend on a single scalar correlation function of the dipole-dipole correlation function. By making a simple medium range oder model, in which the dipole-dipole correlation function is characterized by a single correlation length $k_0^{-1}$ it is possible to make quantitative predictions. It is found that the active temperature depends on the system size, scaling as $L^{4-d}$ at large correlation lengths $L\ll k_0^{-1}$, while in the opposite limit it saturates in three dimensions and diverges logarithmically with the system size in two dimensions.
In three dimensions he velocity correlations decay as $1/r$ for small correlation lengths, while at large correlation lengths the transverse correlation function becomes negative at maximum separation $r\sim L/2$, effect that disappears as the  system increases in size.
\end{abstract}
\maketitle

\section{Introduction}

Microscopic swimmer suspensions constitute an interesting playground for non-equilibrium physics. Energy is continuously taken from the nutrients dissolved in the solution and used to produce directed motion. By the action of living organisms, chemical energy is transformed into kinetic energy in a coherent way, which is then dissipated into heat by viscosity. As an effect of the mutual interaction, swimmers present coherent motion with features similar to turbulence when the suspension is considered as an effective fluid~\cite{Cisneros,Wensink}.

By their motion, swimmers also agitate the fluid and part of the kinetic energy goes to the fluid motion as well.  Near field micro-PIV measurements have helped in explaining the motion of swimmers as well as the perturbations they produce on the fluid~\cite{Drescher2010,Guasto2010}. Experiments  have shown that most of the energy generated by individual bacteria dissipates on the cellular scale and only a small amount is transported to the mesoscale~\cite{Ishikawa}. The fluid agitation has also been indirectly  investigated experimentally following the motion of tracers, which show diffusive behavior at long times~\cite{Wu2000,Leptos09,Mino2010,Valeriani,Lin2011,Zaid2011,Kurtuldu11}. 
When several swimmers are placed in a suspension, they interact by steric forces and also by the  perturbations on the fluid. These hydrodynamic interactions have also an effect of self-induced noise in their dynamics~\cite{Aranson,Saintillan2008,Koch,Evans}. The purpose of this article is to quantify the energy that is stored in the fluid and study how it is spatially structured. 

From a mechanical point of view, swimmers are autonomous objects and, therefore, the total force acting on them vanishes. 
If the swimmers are not isodense, or if another external force acts on them, such as that produced by electric or magnetic fields, the force balance is achieved by exerting a net force on the fluid. In these cases, a swimmer suspension can be considered as a superposition of force monopoles on the fluid. This is similar to what happens in a sedimenting suspension, problem that received considerable attention as primary calculations  indicate that the average fluid velocity would diverge with system size \cite{Smoluchowski}. Properly regularizing the induced and counter flow, it is shown that the mean velocity is finite \cite{Batchelor, BeenakkerMazur} although its variance is predicted to be divergent \cite{CaflishLuke}. Nevertheless, under sedimentation the velocity fluctuations induce unstable density fluctuations that lead to convective flows which regularize the variance. For a review of these results and the subsequent interpretation of the predicted divergent variance see Ref.~\cite{Guazzelli}. 
Here, we focus on the novel effects that appear purely as a result of the swimmer activity. Therefore, we will not consider external forces acting on them.
Consequently, the net force exerted on the fluid vanishes as well and, at first order, swimmers can be modeled as force dipoles. Depending if the dipole is tensile or contractile, the swimmers are classified as pushers or pullers, respectively~\cite{HernandezOrtiz2005, Saintillan2007, Baskaran2009}. In the first category we find bacteria like  \emph{Escherichia coli}  while algae like  \emph{Chlamydomonas reinhardtii}  belong to the second category. The dipolar approximation accurately describes the far field hydrodynamics, but at close distances higher multipoles must be considered as well as the lubrication layer~\citep{Drescher2010, Guasto2010}.
Near field yields, as we show in this article, a regular contribution to the fluid energy while the dipolar part, which we study in detail, gives contributions with long range effects that depend on correlations among swimmers.

The motion of a single swimmer is described by its director $\hat{\mathbf{n}}$, which points along its direction of motion. Axisymmetric swimmers are characterized by a force dipole tensor acting on the fluid
\begin{align}
S_{jk}=\sigma_0 n_in_k,
\end{align}
where $\sigma_0$ is the dipole intensity, negative for pushers and positive for pullers.
The effect of the force dipole on the fluid is obtained by solving the Stokes equations, valid at low Reynolds number. 
The  velocity field produced by a swimmer located at $\mathbf{r}_0$ is
\begin{align}
u_i(\mathbf{r})=J_{ij,k}(\mathbf{r}-\mathbf{r}_0)S_{jk} ,\label{eqn:Umicro}
\end{align}
where $J_{ij,k}$ is the gradient of the Oseen tensor along the direction $k$.  
Summation over repeated indices is assumed.
Its Fourier transform can be easily computed from the Stokes equations
(see, for example, Refs.~\cite{Happel 1965,KimKarilla})
\begin{align}
\widehat{J}_{ij,k}(\mathbf{k})=
\int \dif \mathbf{x}\, e^{-i\mathbf{k}\cdot\mathbf{x}} J_{ij,k}(\mathbf{x})
=\frac{ik_k}{\eta k^2}\left(\delta_{ij}-\frac{k_ik_j}{k^2}\right),
\end{align}
expression that is valid in two or three dimensions. As the swimmer dipole is a symmetric tensor, only the symmetric part of $J_{ij,k}$ is needed, that we call $F_{ijk}=(J_{ij,k}+J_{ik,j})/2$. In Fourier  and real space it reads

\begin{align}
\widehat{F}_{ijk}(\mathbf{k})&=\frac{i}{2 \eta k^2}\left(k_j\delta_{ik}+k_k\delta_{ij}-2\frac{k_ik_jk_k}{k^2}\right),\label{eqn:Fk}\\
F_{ijk}(\mathbf{x}) &=C_d \dfrac{x_i}{x^d}\left(\delta_{jk}-d\dfrac{x_jx_k}{x^2}\right) ,\label{eq.Freal}
\end{align}
where $d$ is the spatial  dimension  that will be either 2 or 3, $C_2=1/(4\pi \eta)$, and $C_3=1/(8\pi \eta)$.
From these expressions it is easily verified that its trace over the last two indices vanishes, indicating that the isotropic part of the force dipole (the pressure) does not contribute to the velocity field, consistent with the hypothesis of incompressibility. Therefore, in what follows we will use, indistinctly the following expressions for the velocity field produced by a swimmer
\begin{align}
u_i(\mathbf{r})=F_{ijk}(\mathbf{r}-\mathbf{r}_0)S_{jk} =F_{ijk}(\mathbf{r}-\mathbf{r}_0)( S_{jk} - S_{ll}\delta_{ik}/d).
\end{align}

When several swimmers are placed in the fluid, by the linearity of the Stokes equations, the resulting flow field is the sum of the effects produced by each swimmer. 
Assuming absence of correlations among swimmers it has been shown that the velocity probability distribution function decays as a power law~\cite{Rushkin2010,Zaid2011,Eckhardt2012}.
The objective of this article is to compute some statistical properties of this flow field considering the effect of correlations among swimmers. Specifically, we will compute the mean square velocity, that can be associated to an {\em active temperature} of the fluid. Also, the equal-time spacial correlations will be computed. It will be shown that, if the system is globally isotropic (that is, the swimmers do not show a collective orientation), both expressions simplify greatly, depending only on a single scalar correlation function of the dipolar tensor. Finally, making simple assumptions on this correlation function the active temperature and velocity correlations are obtained.

\section{Active temperature}

In a suspension  of $N$ swimmers  placed in a volume $V$, the dipolar density is defined as
\begin{align}
s_{jk}(\mathbf{r})&=\sum\limits_{\alpha=1}^N\delta\left(\mathbf{r}-\mathbf{r}^{\alpha}\right)S_{jk}^{\alpha},\label{eqn:Stress}
\end{align}
where $S_{jk}^{\alpha}=\sigma_0 n_j^{\alpha}n_k^{\alpha}$ is the dipolar tensor of the $\alpha$-th swimmer located at $\mathbf{r}^{\alpha}$ and we have assumed that all swimmers have the same dipolar intensity $\sigma_0$. 
In natural  suspensions, swimmers present a distribution of dipolar strengths. Also, for some swimmers like \emph{Chlamydomonas reinhardtii} the dipolar strength oscillates periodically. In Sec.~\ref{sec.variabledipole} we will analyze how the results are extended to these cases.
In terms of this density, the velocity field is 
\begin{align}
u_i(\mathbf{r})&=\int \limits_V \! \dif \mathbf{r}'\, F_{ijk}(\mathbf{r}-\mathbf{r}')s_{jk}(\mathbf{r}').\label{eqn:Vel}
\end{align}

The active temperature, with units of energy, is defined analogously to kinetic theory, proportional to the mean square velocity. 
\begin{align}
T_{act} =& \frac{1}{d}\frac{1}{V}
\int \! \dif \mathbf{r} \, \left\langle u^2(\mathbf{r})\right\rangle \\
=&
\frac{1}{d}\frac{1}{V}\int \! \dif \mathbf{r} \int \! \dif \mathbf{r}_1 \int \! \dif \mathbf{r}_2\, F_{ijk}(\mathbf{r}-\mathbf{r}_1)F_{ilm}(\mathbf{r}-\mathbf{r}_2) \nonumber\\
&  \times \left\langle s_{jk}(\mathbf{r}_1)s_{lm}(\mathbf{r}_2)\right\rangle, \label{eqn:Ta}
\end{align}
where the integral in $\mathbf{r}$  averages in space and $\langle \ldots \rangle$ is an  ensemble average of the possible orientations.

As it is usual when considering the discrete elements that constitute a fluctuating medium, the correlation function of the dipolar density has two contributions. First, there is a self term corresponding to the correlation of a swimmer with itself and there is a cross term given by the correlation of different swimmers. Using Eq. \reff{eqn:Stress}, the correlation function can be decomposed as
\begin{align}
&\langle s_{jk}(\mathbf{r}_1)s_{lm}(\mathbf{r}_2)\rangle  \nonumber \\
&= \left< \delta\left(\mathbf{r}_1-\mathbf{r}_2\right) \sum\limits_{\alpha} \delta\left(\mathbf{r}_1-\mathbf{r}^{\alpha}\right) S_{jk}^{\alpha}S_{lm}^{\alpha} \right>  \nonumber\\ 
&+ \left<\sum\limits_{\alpha\neq\beta} \delta\left(\mathbf{r}_1-\mathbf{r}^{\alpha}\right) \delta\left(\mathbf{r}_2-\mathbf{r}^{\beta}\right) S_{jk}^{\alpha}S_{lm}^{\beta}  \right>  \\
&= \delta\left(\mathbf{r}_1-\mathbf{r}_2\right) A_{jklm} + G_{jklm} \left(\mathbf{r}_1-\mathbf{r}_2\right) ,\label{eq.defG}
\end{align}
expression that defines the constant tensor $A$ and the correlation tensor $G$. In the previous expression it has been assumed that the system is spatially homogeneous and therefore the correlation tensor only depends on the relative distance. Substituting \reff{eq.defG} into \reff{eqn:Ta} gives the active temperature in terms of the statistical properties of the swimmers. The first term gives a contribution that depends on the properties of individual swimmers while the second term depends on their correlations. We analyze  both contributions separately.

\subsection{Self contribution}
As seen from \reff{eq.defG} the self part deals with the effect of individual swimmers. Considering the homogeneity of the suspension,  this contribution to the active temperature can be computed multiplying by $N$ the effect of a single swimmer. That is,
\begin{align}
T^{\mathrm{self}}_{\mathrm{act}} &= \frac{N}{dV} \int \! \dif \mathbf{r} \,  \left[F_{ijk}(\mathbf{r}) \sigma_0 n_j n_k \right]^2 \nonumber\\
&= \frac{\rho\sigma_0^2 C_d^2}{d}  \int \! \dif \mathbf{r} \,  \frac{1}{r^{2d-2}} 
\left[1-d\left(\frac{\mathbf{r}\cdot\mathbf{n}}{r} \right)^2 \right]^2,
\end{align}
where we have defined the swimmer number density $\rho=N/V$ and we used the expression  \reff{eq.Freal} for $F$.
The resulting expression presents a divergence at short distances originated in neglecting the finite size of the swimmer. We introduce a finite size cutoff $a$ comparable to the swimmer size. At this scale, the excluded volume must be considered and other multipoles should be included as well in the description of the flow field~\cite{Drescher2010, Guasto2010}. In two dimensions, the integral also diverges at long distances. We introduce a large scale cutoff $L$ equal to the system size.
After these considerations, the result is
\beq
T^{\mathrm{self}}_{\mathrm{act}} = \frac{\rho\sigma_0^2}{32\pi\eta^2}  \log(L/a)
\eeq
in two dimensions and
\beq
T^{\mathrm{self}}_{\mathrm{act}} = \frac{\rho\sigma_0^2}{60\pi\eta^2}  \frac{1}{a}
\eeq
in three dimensions. Note that in both cases, the active temperature is dominated by the short scale and therefore a precise modeling of the near flows is essential to quantitatively characterize the induced agitation on the fluid.

Higher multipolar terms produce contributions to the velocity field that decay as $1/r^d$ or faster away from the swimmer. These terms give finite, size-independent, contributions to $T^{\mathrm{self}}_{\mathrm{act}}$ and therefore can be absorbed into $a$. The same conclusion can be drawn in the analysis on the contribution of the higher multipoles in the correlation part of the active temperature. Hence, in the previous expressions the parameter $a$ takes into account the swimmer size and the near field contributions.

\subsection{Correlation contribution}
Using the spacial homogeneity of the correlation tensor, its contribution to the active temperature can be written as
\begin{align}
T^{\mathrm{corr}}_{\mathrm{act}} &= \frac{1}{d}\int \! \dif \mathbf{x} \, H_{jklm}(\mathbf{x})G_{ijkl}(\mathbf{x}) ,\label{eqn:Tc2}
\end{align}
where
$H_{jklm}(\mathbf{x}) = \int \! \dif \mathbf{y} \, F_{ijk}(\mathbf{y})F_{ilm}(\mathbf{y}-\mathbf{x})$, that in Fourier space satisfies $\widehat{H}_{jklm} = \widehat{F}^*_{ijk}\widehat{F}_{ilm}$. 
Using \reff{eqn:Fk} it can be written explicitly 
\begin{align}
\widehat{H}_{jklm}=&\frac{1}{4 \eta^2 k^2} \left[-4\frac{k_jk_kk_lk_m}{k^4}\right. \nonumber\\
&+\left. \frac{1}{k^2}\left(k_jk_l\delta_{km}+k_jk_m\delta_{kl}+k_kk_l\delta_{jm}+k_kk_m\delta_{jl}\right)\right]. \label{eqn:H}
\end{align}

The correlation tensor \reff{eq.defG} is symmetric in the each  pair of indices $jk$ and $lm$ as well as under the interchange of the pair $jk$ with $lm$. 
Assuming isotropy, its Fourier transform can be written in the most general form as
\begin{widetext}
\begin{align}
\widehat{G}_{jklm}(\mathbf{k}) =& \widehat{G}_{1}(k)\delta_{jk}\delta_{lm}
+\frac{\widehat{G}_{2}(k)}{k^2}\left(\delta_{jk}k_lk_m+\delta_{lm}k_jk_k\right)+\widehat{G}_{3}(k)\left(\delta_{jk}\delta_{lm}+\delta_{jl}\delta_{km}+\delta_{jm}\delta_{kl}\right) \nonumber \\
&+\frac{\widehat{G}_{4}(k)}{k^4}k_jk_kk_lk_m
 +\frac{\widehat{G}_{5}(k)}{k^2}\left(\delta_{jk}k_lk_m+\delta_{lm}k_jk_k+ \delta_{jl}k_kk_m+\delta_{jm}k_kk_l +\delta_{kl}k_jk_m+\delta_{km}k_jk_l\right) \label{eq.Gtensor}
 \end{align}
\end{widetext}
in terms of a small number of scalar functions multiplied by some specific tensors. 

Once \reff{eq.Gtensor} and \reff{eqn:H} are substituted in the expression for the active temperature \reff{eqn:Tc2} there are important algebraic simplifications and we obtain the compact expression
\begin{align}
T^{\mathrm{corr}}_{\mathrm{act}}  &=\frac{d-1}{d(2\pi )^d\eta^2} \int \! \dif \mathbf{k} \, \frac{\widehat{G}_{0}(k)}{k^2} ,\label{eqn:Tc3}
\end{align}
where $\widehat{G}_{0}=\widehat{G}_{3}+\widehat{G}_{5}$.  By dimensional analysis the Fourier transform of this function can be expressed as $G_0(r) = \sigma_0^2\rho^2 g(r)$, where $g$ is dimensionless. At  long distances swimmers are uncorrelated and $g\to1$ while at short distances $g\to0$ due to the excluded volume. The long distance value of $g$ implies a Dirac delta term in $\widehat{G}_0$. However, as usual when theories with long range interactions are homogenized~\cite{Torquato,Kanuan}, this $\delta$ term does not contribute as can be directly checked by integrating a constant value in \reff{eqn:Tc2} in a finite volume and then taking the volume to infinity. Therefore, the $\delta$ term in $\widehat{G}_0$ will not be considered.

Note that $\widehat{G}_0$ can be measured in experiments or computed in simulations of discrete elements by directly computing the full contraction $\widehat{G}_{jklm} \widehat{H}_{jklm}$ of the dipole-dipole structure factor.

\subsection{Model with medium range order}
The statistical  properties of the fluid velocity depend on $\widehat{G}_0$, which is a scalar function that characterizes the dipole-dipole tensor correlations. To our knowledge, this function has not been measured in swimmer suspensions. The simplest assumption we can make is that it is characterized by a single correlation length $k_0^{-1}$ that, eventually, can diverge at a critical point as it could happen  in a swarming phase or in other phases with collective order~\cite{TonerTu}.  We  consider, therefore, a Lorentzian model with medium range order
\beq
\widehat{G}_0(k) =  \frac{\Gamma\sigma_0^2}{k^2+k_0^2}, \label{eq.modelG0}
\eeq
where $\Gamma$ is a measure of the correlation intensity and we have factored out the dependence on the dipole strength. 
We will use this model to analyze the behavior of the active temperature and velocity correlations 
as it captures the general properties of systems with medium range order for the dipolar parameter. More accurate models, obtained from experiments, discrete element simulations~\cite{Wensink} or continuous  models~\cite{rheology} will change only qualitatively the picture below if they are characterized by a single correlation length. More complex  models, with different scaling at large distances, should be worked out separately.

\begin{figure}[htb]
\begin{center}
\includegraphics[clip,width=.9\columnwidth]{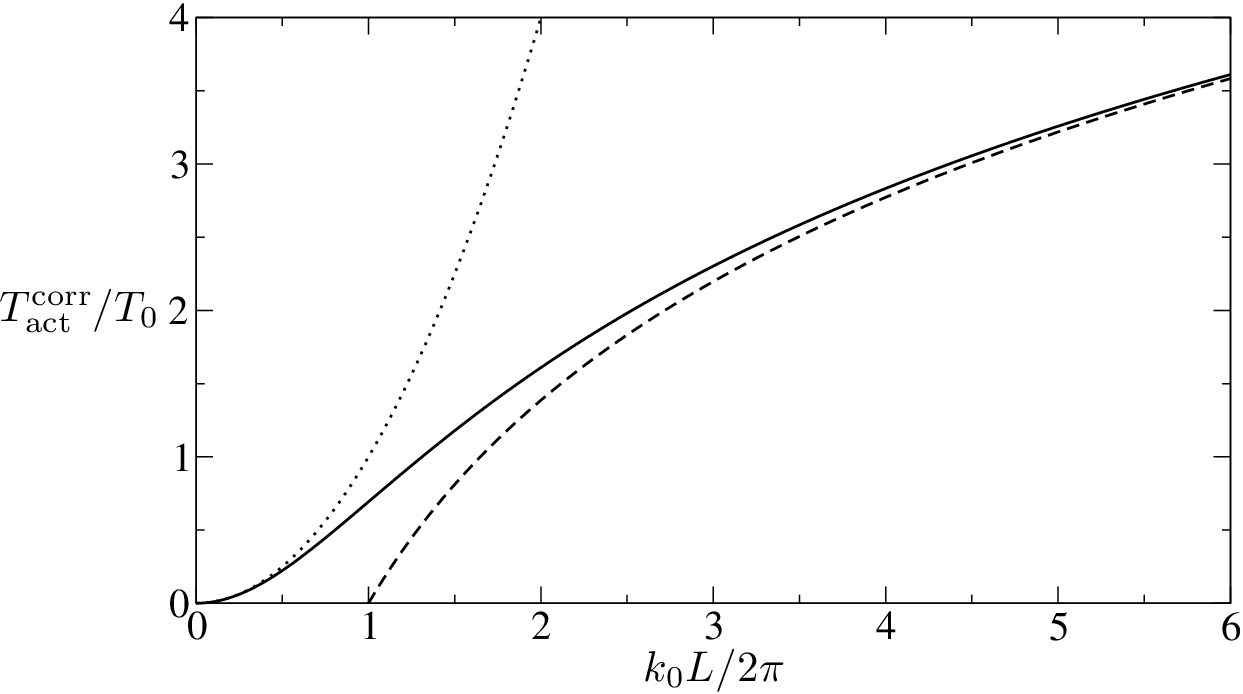}
\includegraphics[clip,width=.9\columnwidth]{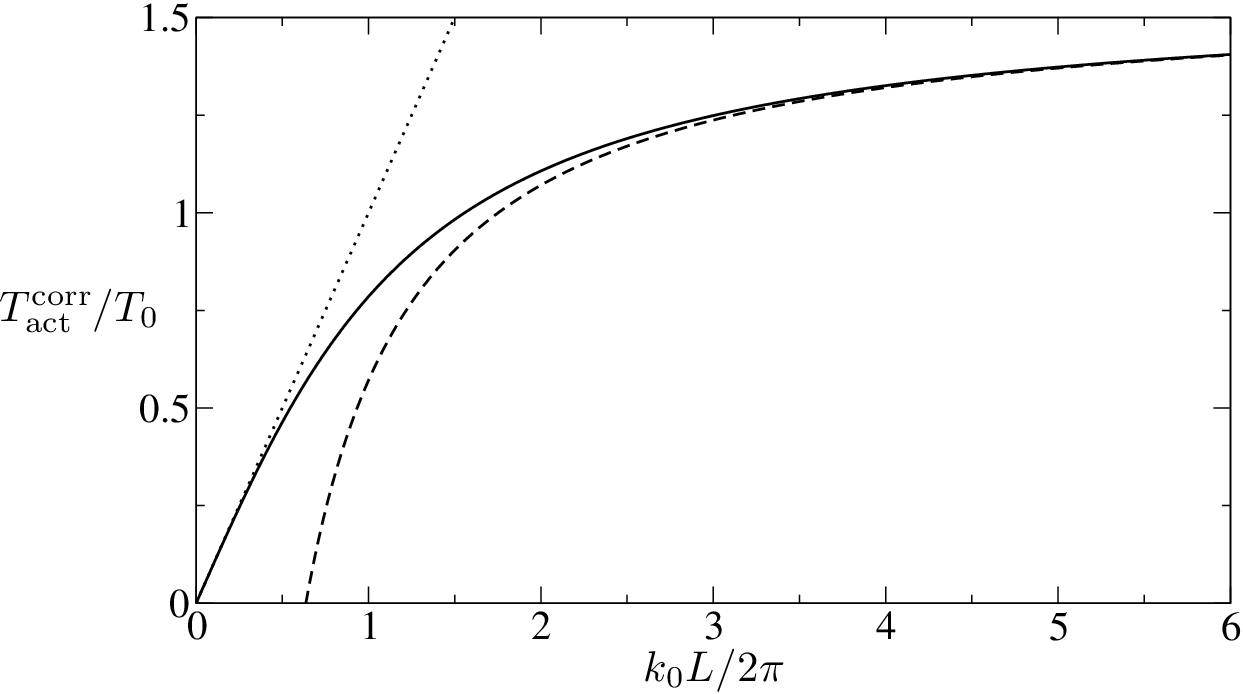}
\end{center}
\caption{Correlation contribution to the active temperature as a function of the normalized system size $k_0L/2\pi$. The temperatures are normalized by $T_0=\frac{\Gamma\sigma_0^2}{8\pi\mathrm{k_B}\eta^2k_0^2}$ is two dimensions (top) and by $T_0=\frac{\Gamma\sigma_0^2}{3\pi^2\mathrm{k_B}\eta^2k_0}$ in three dimensions (bottom). The solid curve is the full expression, while the dashed and dotted curves are the long and small system size approximations, respectively.}
\label{fig:Tc}
\end{figure}

With the medium range  model \reff{eq.modelG0} the correlation part of the active temperature is
\beq
T^{\mathrm{corr}}_{\mathrm{act}}  =\frac{\Gamma\sigma_0^2}{8\pi\eta^2k_0^2} 
\log\left[1+\left(\frac{k_0L}{2\pi}\right)^2\right]
\eeq
in two dimensions and
\beq
T^{\mathrm{corr}}_{\mathrm{act}}  =\frac{\Gamma\sigma_0^2}{3\pi^2\eta^2k_0} \arctan\left(\frac{k_0L}{2\pi}\right) 
\eeq
in three dimensions, which are shown in Fig. \ref{fig:Tc}.

When the correlation length  is much larger compared to the system size, $k_0L\ll1$, as it can happen close to a critical point when the correlation length diverges, the active temperature shows important finite size effects. In this regime it goes as $T^{\mathrm{corr}}_{\mathrm{act}}\sim L^{4-d}$. 

In the opposite limit of short correlation lengths $k_0L\gg 1$ the behavior in two and three dimensions is different. In three dimensions, the active temperature goes to a finite value that depends on the intensity $\Gamma$ of the correlations. In two dimensions, there is a logarithmic dependence with the system size that, although is weaker than the previously found in the small correlation length case, can still be observable. Note that in this regime, in two and three dimensions, the size dependence has the same scaling as $T^{\mathrm{self}}_{\mathrm{act}}$ and the relative intensities of both depend on the degree of correlation between swimmers.

\section{Spatial velocity correlations}
To complement the information provided by the active temperature, we consider the velocity correlation function to characterize the spatial structure of the flow
\begin{align}
C_{ij}(\mathbf{x}) &=\frac{1}{V}\int \! \dif \mathbf{r} \, \left\langle u_i(\mathbf{r})u_j(\mathbf{r}+\mathbf{x})\right\rangle .\label{eqn:Defgik}
\end{align}
Using the isotropy of the system, this tensor can be decomposed into the transverse and longitudinal parts as
\begin{align}
C_{ij}(\mathbf{x}) &= C_{\perp}(x)\left(\delta_{ij}-\frac{x_ix_j}{x^2}\right)+C_{\parallel}(x)\frac{x_ix_j}{x^2}, \label{eqn:Descomp}
\end{align}
where now $C_{\perp}$ and $C_{\parallel}$ are scalar functions of the distance. If vortex-like structures developed as in turbulent flows, the  transverse part would present a negative region at the vortex characteristic size.

It is direct to obtain a Fourier representation of the correlation function in terms of the dipole-dipole tensor correlation function $G$
\begin{align}
\widehat{C}_{ij}(\mathbf{k}) &=  \widehat{G}_{klmn}(\mathbf{k})\widehat{H}_{ijklmn}(-\mathbf{k}), \label{eqn:gik}
\end{align}
where $\widehat{H}_{ijklmn}(\mathbf{k}) = \widehat{F}_{ikl}^*(\mathbf{k})\widehat{F}_{jmn}(\mathbf{k})$ is given explicitly by
\begin{widetext}
\begin{align}
\widehat{H}_{ijklmn}(\mathbf{k}) =& \frac{1}{4\eta^2 k^4} \bigg[\left(k_kk_m\delta_{jn}\delta_{il}+k_kk_n\delta_{jm}\delta_{il}+k_lk_m\delta_{jn}\delta_{ik}+k_lk_n\delta_{jm}\delta_{ik}\right) \nonumber \\
&  \left. -\frac{2}{k^2}\left(k_kk_jk_mk_n \delta_{il}+ k_lk_jk_mk_n \delta_{ik}+k_mk_ik_kk_l\delta_{jn}+k_nk_ik_kk_l\delta_{jm}\right)+4\frac{k_ik_kk_lk_jk_mk_n}{k^4}\right].
\end{align}
\end{widetext}

Although it looks involved, once this expression and the tensorial decomposition of $\widehat{G}$ are replaced in \reff{eqn:gik}, algebraic simplifications take place and 
we get the extraordinary simple result
\begin{equation}
\widehat{C}_{ij}(\mathbf{k}) 
= \frac{\widehat{G}_{0}(k)}{\eta^2k^2}\left(\delta_{ij}-\frac{k_ik_j}{k^2}\right).\label{eqn:gik2}
\end{equation}
Again, this expression only depends on $\widehat{G}_0$. Then, measuring this correlation function or by modelling it we can get detailed information on the flow structure.

We note that in Fourier space the velocity correlations only have transverse component, consistent with the incompressibility condition that is expressed as $\widehat{C}_{ij}k_j=0$. However, when going back to real space both longitudinal and transversal components appear, given by
\begin{align}
C_{\perp}(x) =&  \frac{(2\pi)^{-d}}{(d-1) }\int \! \dif \mathbf{k} \, e^{i\mathbf{k}\cdot \mathbf{x}}\frac{\widehat{G}_{0}}{\eta^2k^2}
\left(d-2+\frac{(\mathbf{k}\cdot \mathbf{x})^2}{k^2x^2}\right), \label{eqn:I1} \\
C_{\parallel}(x) =& (2\pi)^{-d}\int \! \dif \mathbf{k} \, e^{i\mathbf{k}\cdot \mathbf{x}}\frac{\widehat{G}_{0}}{\eta^2k^2}\left(1-\frac{(\mathbf{k}\cdot \mathbf{x})^2}{k^2x^2}\right). \label{eqn:I2}
\end{align}

\subsection{Model with medium range order}
The transverse and longitudinal correlation functions can be computed using the medium range order model \reff{eq.modelG0}. Here, we will present results only for the three dimensional case as in two dimensions the results show size effects that mask the spacial dependence. 
Even in three dimensions, special consideration must be used as the limit of infinitely long system does not commute with the limit of long correlation lengths. 

\begin{figure}[htb]
\begin{center}
\includegraphics[width=.9\columnwidth]{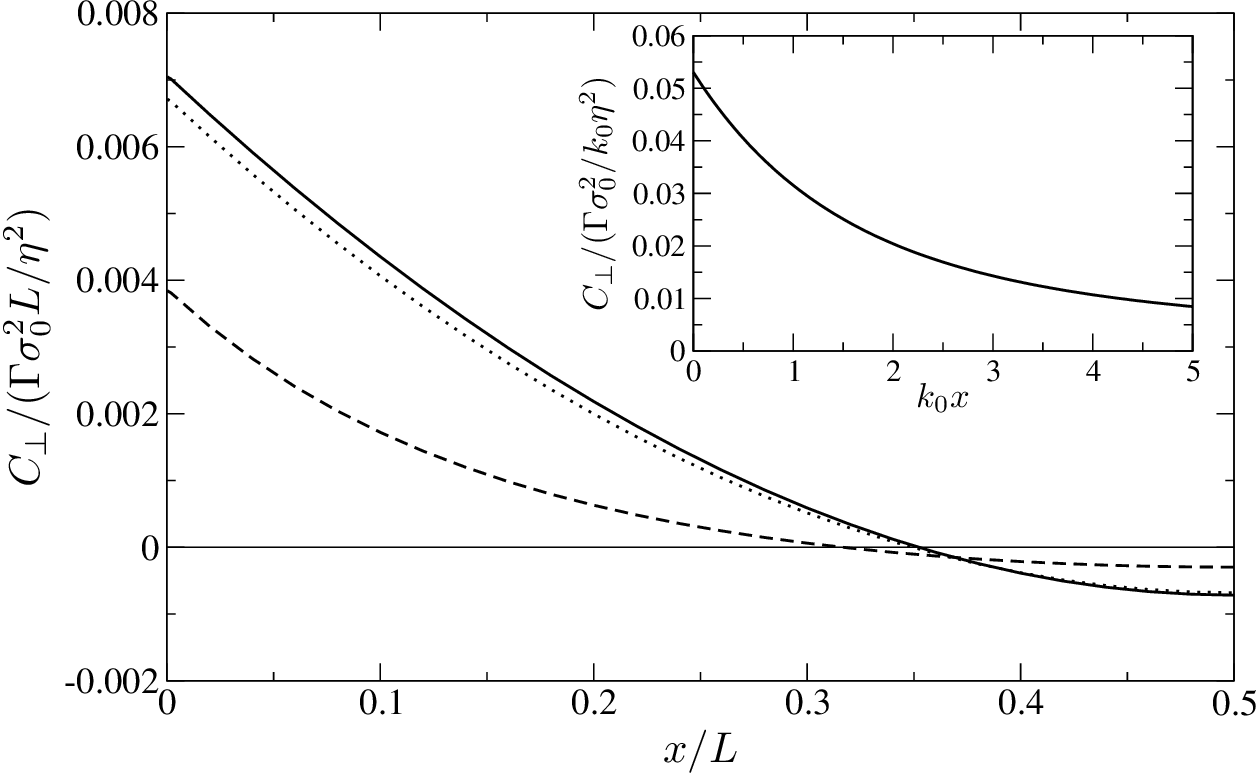}
\includegraphics[width=.9\columnwidth]{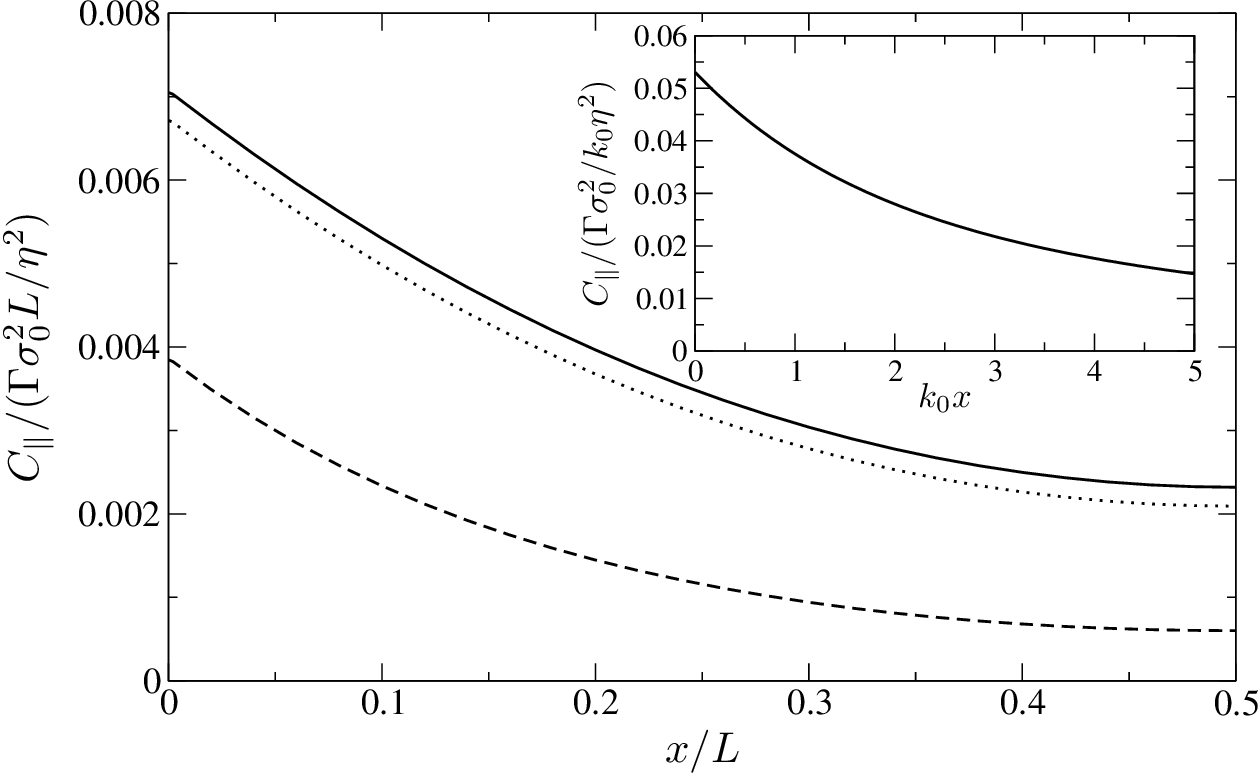}
\end{center}
\caption{Transverse (top) and longitudinal (bottom) velocity correlations as a function of the point separation $x$ for three correlation lengths [$k_0L=0.1$ (solid line), $k_0L=2.0$ (dotted line) and $k_0L=10.0$ (dashed line)]. The inset shows the case of $k_0L\to\infty$. In the main figures the distances are normalized to the system size $L$ and in the inset to the correlation length $k_0^{-1}$. }
\label{fig:corr}
\end{figure}

We first consider the limiting case $k_0L\gg1$ when the  integrals in $\mathbf{k}$ are unbounded and can be computed analytically, resulting in
\begin{align}
C_{\perp}(x) =& \frac{\Gamma\sigma_0^2}{4\pi\eta^2 k_0} \left[\frac{1}{2\xi} +\frac{1}{\xi^3} - e^{-\xi} \left(\frac{1}{\xi}+\frac{1}{\xi^2}+\frac{1}{\xi^3} \right)   \right], \label{eq.cperpinf}\\
C_{\parallel}(x) =& \frac{\Gamma\sigma_0^2}{2\pi\eta^2 k_0} \left[\frac{1}{2\xi} -\frac{1}{\xi^3} + e^{-\xi} \left(\frac{1}{\xi^2}+\frac{1}{\xi^3} \right)   \right]  .\label{eq.cparinf}
\end{align}
where $\xi=k_0r$. We note that both expressions are finite at short distances and at large distances decay as $1/r$. Note that long range velocity correlations develop despite the fact that the  orientational correlation of swimmers is of medium range only. Also, both correlation functions are always positive and therefore the system does not develop vortex-like structures (see insets of Fig. \ref{fig:corr}).

For finite system we could proceed by imposing a lower bound $2\pi/L$ to the wavevectors as it was done when computing the active temperature. However, to capture more precisely the spatial structure we proceed instead by doing a discrete sum over Fourier modes $\mathbf{k} = 2\pi(n_x,n_y,n_z)/L$, excluding the $\mathbf{k}=0$ term as it was explained before. 
The sum rapidly converges when $k_0L$ is small, needing only few terms, while when $k_0L$ gets large  an increasing number of terms must be considered to make the sum converge. But in this later case, we can use instead the asymptotic expressions \reff{eq.cperpinf} and \reff{eq.cparinf}. The results for different values of $k_0L$ are presented in Fig. \ref{fig:corr}. 
When $L$ is finite, the correlation functions are even in $x$ and periodic in $L$ and therefore, only the region $0\leq x\leq L/2$ is presented.

When the distance is normalized to the system length, the curves collapse for small box lengths up to $k_0L\sim 1$ while for larger sizes the curves separate. The longitudinal velocity correlation is always positive while  the transversal correlation function becomes null at $x/L\sim0.35$, presenting a negative region from this point up to the maximum separation at $x/L=0.5$. The amplitude of the negative anticorrelation is small. It becomes smaller when $k_0L$ increases and, finally, it disappears in the limit $k_0L\to\infty$. As the distance at which it becomes negative scales directly with the system size, we would not call it a  vortex-like structure, but rather a finite size effect phenomenon. 

\section{Dipole strength variability and oscillatory dipoles}\label{sec.variabledipole}

In the previous sections we considered that all swimmers have the same dipolar strength $\sigma_0$. In natural swimmer suspensions there is normally a distribution of intensities. It can be directly verified that, if this is the case, our previous results remain valid except that everywhere $\sigma_0^2$ should be replaced by $\langle \sigma^2\rangle$, where the average is taken over the swimmer distribution. 

Also, some swimmers like \emph{Chlamydomonas reinhardtii} propel via the periodic motion of a pair of flagella. The complex flow results from a considering several terms in the multipolar expansion. However,  as mentioned before, the eventual size effects and velocity correlations are only due to the dipolar contribution. In this case, the dipolar strength varies periodically. To show how our expressions are modified we model this variability as $\sigma=\sigma_A+\sigma_B \cos \phi$, where the phase $\phi$ is assumed to vary linearly with time. The self-contribution to the active temperature is obtained by simply replacing $\sigma_0^2$ by the average $\langle \sigma^2\rangle=\sigma_A^2+\sigma_B^2/2$. The dipole-dipole correlation contribution to the active temperature and velocity correlations depends on the phase correlation between swimmers $\langle \sigma_1\sigma_2\rangle = 
\sigma_A^2 + \sigma_B^2 \langle \cos(\phi_1-\phi_2)\rangle/2$. 
If swimmers show no phase correlation, only the average strength intensity is relevant, while if swimmers show synchronization as for example in Ref. \cite{Uchida} then the temporal modulation of the dipolar strength enters into play. Phase synchronization will show a correlation length that will normally be different from the orientation correlation length considered in the previous sections. It is sensible to model in this case the total dipole-dipole correlation function as a sum of two Lorentzian 
\beq
\widehat{G}_0(k) = \frac{\Gamma_A \sigma_A^2}{k^2+k_A^2} + \frac{\Gamma_B \sigma_B^2}{k^2+k_B^2}. 
\eeq
From this model, the same analysis as those done in the previous sections can be performed with similar conclusions, although the two correlation length scales and the relative intensities should be considered.

\section{Conclusions}
The fluid agitation produced by an isotropic and homogeneous swimmer suspension is investigated in which swimmers are modeled as force dipoles.
Appealing to general properties of isotropy and homogeneity of the Stokes equations it is found that generically, the equal-time statistical properties depend on a single correlation function of the dipole-dipole correlation function that characterizes the orientational correlation of swimmers. In discrete simulations~\cite{Wensink} this function can be directly computed.  Experiments that measure the swimmer orientation would be desirable, but limitations of visualizing individual swimmers exist in the bulk of dense suspensions, while it has been possible to track position and orientation in dilute bulk suspensions \cite{Liao07} or  near surfaces in quasi two dimensional geometries \cite{Zhang10}. To our knowledge no measurement of the dipole-dipole correlation have been made. Finally, continuous models need to include the dynamics of the dipolar tensor as in Ref.~\cite{rheology} as hydrodynamic-like models for the density and velocity fields only provide insufficient information. 

By making a simple model of this correlation function, based on a single correlation length, it is shown that the active temperature presents strong size effects that should be noticeable in experiments, for example, by measuring the induced diffusion on tracers or by micro particle image velocimetry. If confining walls are used, they induce a faster decay of the velocity fields and the size effects could not be observed near walls, but in the bulk they could still be observable.
The velocity correlations do not show the appearance of coherent structures, except in finite systems, in which the transverse component is negative for points separated by half of the system size, but these cannot be homologated to vortex-like structures. 

\section*{Acknowledgment}
This research is supported by Fondecyt Grant 1100100 and Project Anillo ACT 127. C.P.-R. acknowledges the support of a CONICYT grant.

\end{document}